\documentclass[grl]{agutex}  

\usepackage{lineno}

\usepackage{amsbsy}

\usepackage{amsmath}
\usepackage{amsthm}
\usepackage{eucal}
\usepackage{amssymb}
\usepackage{mathrsfs}

\usepackage{color}

\usepackage[pdftex]{graphicx}
\usepackage{graphics}
\DeclareGraphicsExtensions{.pdf,.PDF,.jpg,.JPG}
\graphicspath{{/Users/patou/Documents/RAPPORT_BOULOT/NANO-POUSSIERES/FIGURES/}}

\authorrunninghead{SCHIPPERS ET AL.}

\titlerunninghead{Nanodust near 1 AU measured by Cassini}  

\authoraddr{P. Schippers,
LESIA - CNRS - Observatoire de Paris, 5 place Jules Janssen, 92195 Meudon, France.
(patricia.schippers@obspm.fr)}

\begin{document}

%
%
%
\setkeys{Gin}{draft=false}
%

%

%
%


\title{Nanodust detection near 1 AU from spectral analysis of Cassini/RPWS radio data}


%
%


\authors{P. Schippers\altaffilmark{1},
N. Meyer-Vernet\altaffilmark{1},
A. Lecacheux\altaffilmark{1},
W.S. Kurth\altaffilmark{2},
D. G. Mitchell\altaffilmark{3},
N. Andr\'e\altaffilmark{4}
}

\altaffiltext{1}
{LESIA - CNRS - Observatoire de Paris, 5 place Jules Janssen, 92195 Meudon, France.}
\altaffiltext{2}
{Department of Physics and Astronomy, University of Iowa, Iowa City, Iowa, USA.}
\altaffiltext{3}
{ Applied Physics Laboratory, John Hopkins University, Laurel, Maryland, USA.}
\altaffiltext{4}
{ IRAP, 9 Avenue du Colonel Roche, 31028 Toulouse, France.}

%
%

\begin{abstract}

{Nanodust grains of a few nanometer in size are produced near the Sun by collisional break-up of larger grains and picked-up by the magnetized solar wind. They have so far been detected at 1 AU by only the two STEREO spacecraft.} Here we analyze the spectra measured by the radio and plasma wave instrument onboard Cassini during the cruise phase close to Earth orbit; they exhibit bursty signatures similar to those observed by the same instrument in association to  nanodust stream impacts on Cassini near Jupiter. The  observed wave level and spectral shape reveal impacts of nanoparticles at about $ 300$ km/s, with an average flux  compatible with that observed by the radio and plasma wave instrument onboard STEREO and with the interplanetary flux models.

\end{abstract}

\begin{article}

\section{Introduction}

Nanodust, which lie at the low end of the mass distribution of solar system bodies, bridge the gap between molecules {and sub-micron} grains. {Their large surface-to-volume ratio implies that most  of the atoms lie at the grains' surface, favoring the interactions with the ambient medium}, whereas  their high charge-to-mass ratio enables them to be picked-up by moving magnetized plasmas  similarly {to ions}.

{Nanodust grains have been observed in situ in a number of environments, {such as those} of comets \citep{Utterback1990}, Jupiter \citep{Grun1993,Zook1996}, and Saturn  \citep{Kempf2005,Hsu2012}. In the interplanetary medium, the presence of nanodust produced in the inner solar system by dust collisional fragmentation {\citep[e.g.,][]{Grun1985}} and accelerated by the magnetized solar wind to about 300 km/s was predicted by \citet{Mann2007} and discovered aboard the \textit{Solar Terrestrial Relations Observatory} (STEREO) \citep{Driesman2008} at 1 AU \citep{MeyerVernet2009b}. The presence of these particles has been suggested to have major consequences, especially in the inner solar system where their trapping  \citep{Czechowski2012} may produce high nanodust densities  possibly contributing to the diffuse X-ray background  \citep{Kharchenko2012}, whereas their contribution to the inner source of solar wind pick-up ions (e.g. \citep{Gloeckler1998}) is a disputed question \citep{Wimmer-Schweingruber2010,Mann2012}.}
{The presence of nanodust in solar system plasmas is therefore of major consequence.}

Even though the STEREO preliminary observations, which used the low-frequency receiver of the STEREO/WAVES instrument \citep{Bougeret2008}, have been subsequently confirmed and improved by detailed analyses of data from an {independent receiver, the time-domain sampler  \citep{Zaslavsky2012}}, and by five years of  data \citep{LeChat2013}, a confirmation onboard other spacecraft would be of major importance. This is the subject of the present letter. We analyze the power spectral density measured with {the radio and plasma wave science (RPWS) instrument onboard Cassini} in the solar wind, just after the Earth fly-by (August 18, 1999), when the instrument was turned-on for one month. We show that these data reveal impacts on the spacecraft of nanodust streams accelerated by the solar wind, with average properties compatible with STEREO/WAVES observations.

{\section{\label{sec2} Wave instruments as dust detectors}}

Wave instruments are routinely used  in space to measure in situ plasma properties via quasi-thermal noise spectroscopy with great accuracy and an equivalent cross-section of detection much larger than traditional detectors \citep{MeyerVernet1998a}. This technique is based on the electric voltage pulses induced on electric antennas by passing or impacting plasma particles \citep{MeyerVernet1989}. Similarly, impact ionization produced by  dust grains impinging on  the spacecraft or antennas at high speed produces electric pulses that can be detected by wave instruments. This technique, pioneered when the Voyager spacecraft crossed Saturn's dusty rings \citep{Aubier1983,Gurnett1983}, has been  refined (e.g. \citep{MeyerVernet1996,MeyerVernet1998b}). {With Cassini/RPWS,  it has been used to measure fast Jovian nanodust streams \citep{MeyerVernet2009a} simultaneously with the \textit{Cosmic dust analyzer} (CDA) instrument  \citep{Srama2004,Kempf2005} and it is currently used to measure microdust in Saturn's magnetosphere \citep{Moncuquet2013,Ye2013}.}

{Wave instruments are complementary to dedicated dust detectors \citep{Auer2001} because of their high sensitivity due to the effective surface area for dust detection being the whole spacecraft surface, with a $2 \pi$ steradians  acceptance solid angle. This is at the expense of a lack of velocity determination and large uncertainties.} 
{However, this technique requires generally two conditions. {Since it is based on the detection of voltage pulses produced by the electric charges released by impact ionization, it first requires that the impact speed exceeds a few km/s \citep{Auer2001}.
Since the main target is the spacecraft surface, which tends to recollect the impact charges of sign opposite to that of its floating potential (i.e., the electrons in the solar wind), voltage levels measured between each antenna arm and the spacecraft (the so-called monopoles) are expected to be similar. It then also requires that the antennas must be connected in monopole mode instead of dipole mode ( i.e. measurement of the difference of potential between two antenna arms).}
Unfortunately, most wave instruments use antennas in dipole mode. This is so because this mode is generally better adapted for wave measurements (yielding a greater and better determined antenna effective length), except for direction finding or polarization determination purposes. In that dipole mode, the dust impact signal is generally much smaller, since a change in spacecraft potential modifies negligibly the antenna potential, so that the detection requires asymmetries of the system  \citep{MeyerVernet2014}. This holds except when the constraints imposed by mission priorities impose less adequate  antenna mounting geometries as was the case for STEREO \citep{Driesman2008}, for which the antenna location caused them to be frequently immersed in the impact plasma \citep{Zaslavsky2012}.} 

{The difference between monopole and dipole mode for dust detection is well illustrated with Cassini/RPWS, which includes three electric antenna booms,  one of which is always used as a monopole while the two others are either used as a dipole (i.e. the "dipole+monopole" mode) or a monopole (i.e. the "three monopole" mode or commonly called "Direction Finding" mode). Impacts of E ring microdust  \citep{Moncuquet2013} and Jovian nanodust \citep{MeyerVernet2009a} were observed to produce similar signals on each monopole and much smaller signals on the dipole \citep{MeyerVernet2014}, in contrast to radio and plasma waves which generally produce different voltages on the different monopoles and greater voltages in  dipole mode.} 

{A further property of the voltage pulses produced by dust impacts is that the pulse rise time $\tau _r $ is of the order of tens of $\mu$s. Therefore the voltage power spectrum has a steep spectral shape, proportional to $f^{-4}$ at frequencies exceeding $1/2\pi \tau_r$, of order a few kHz.}

{\section{Dust impact identification}}

{During the investigated time period, the Cassini spacecraft was oriented such that the high gain antenna (HGA) was pointing to the sunward direction and the Huygens probe was put in the spacecraft velocity plane \citep{Matson2002}.  The three antenna booms of the RPWS instrument are connected to a suite of five receivers designed to measure electrostatic and electromagnetic waves in Saturn's magnetosphere.} In the present study, we analyze the data from the high frequency receiver HFR (3.5 kHz-16MHz) from August 20 (day 232) to September 15 (day 258) of 1999. 
 Figure \ref{RPWS_Spectro}a displays the time-frequency electric power spectral density measured by the RPWS/HFR between September 6 and 10 in monopole mode. In addition to solar type III bursts and plasma quasi-thermal noise around the electron plasma frequency ($f_{pe}$), we observe low frequency bursty activity characterized by sudden voltage enhancements, below 20 kHz.
  
 Figure \ref{RPWS_Spectro}b displays two spectral cuts on September 8 (day 251), one at 20:19 UT  (black solid line) and the other at 20:20 UT  (black dashed line).
 The spectrum at 20:20 UT displays a trend close to a power law in $\omega^{-2}$ (blue curve in Figure \ref{RPWS_Spectro}). This signature is typically produced by the shot noise due to impacts of the ambient electrons on the spacecraft body and antennas \citep{MeyerVernet1983,MeyerVernet1989}. 
The spectrum at 20:19 UT displays a much higher amplitude with a steeper slope, proportional to a power law $\omega^{-|n>{2}|}$, with index $n$ close to 4 . Such a spectral shape is similar to the dust particle impact signature \citep{MeyerVernet1986}, and routinely observed on Cassini/RPWS in association to nano \citep{MeyerVernet2009a} and micro \citep{Moncuquet2013} dust impacts. Indeed, as indicated in Section  \ref{sec2}, the impact of a dust grain on the spacecraft creates a plasma cloud whose expansion decouples the ion and electrons which are then recollected by the spacecraft surface, depending on the sign of its electric potential. The Fourier transform of the approximate signal induced by one dust particle impact is \citep{Aubier1983}: 
 \begin{linenomath*}
 \begin{equation}
 V_{fi}^2 \simeq\ 2 \delta V^2 \omega^{-2}(1+\omega^2\tau_r^2)^{-1}
 \label{Eqn_Dust1}
 \end{equation}
 \end{linenomath*}
 where $\delta$V is the maximum pulse amplitude, $\tau_r$ is the pulse rise time, at frequencies $f=\omega/2\pi$ much higher than the inverse of the pulse's decay time $\tau_d$.  When $\omega \tau_r \gg 1$, Equation (\ref{Eqn_Dust1}) simplifies into 
  \begin{linenomath*}
 \begin{equation}
 V_{fi}^2\simeq 2 (\delta V^2/\tau_r^2)\omega^{-4}\
 \label{Eqn_Dust2}
 \end{equation}
  \end{linenomath*}
 
Models (\ref{Eqn_Dust1}) and (\ref{Eqn_Dust2}) have been adjusted to the 20:19 UT spectrum and are displayed in red and green, respectively, in Figure \ref{RPWS_Spectro}b.
Model (\ref{Eqn_Dust1}) reliably fits the measurement, with $\tau_r \simeq$ 30 $\mu s$.

\section{Dust flux determination}

\subsection{MIMI INCA discharges}
To characterize extensively the dust flux profile during the time period of interest, we have to take into account an issue reported by the \textit{Magnetospheric Imaging Instrument} (MIMI) \citep{Krimigis2004} team concerning collimator discharges during the early mission. It was reported that with the negative high voltage applied to the \textit{Imaging Neutral Camera} (INCA) \citep{Krimigis2004} charged particle rejection plates turned to maximum values, discharges were experienced that disturbed the measurement of the MIMI instrument itself and might have contaminated the radio measurements of RPWS. The origin of such discharges has not been fully elucidated yet but sunlight shining on the plates and dust hits appear to be responsible for some of them. In order not to use contaminated data, we had to identify and discard the discharge events in our dataset. To do so, we used noise indicators observable in a few MIMI/INCA control parameters (the instrument incorporated a discharge monitor for just this possibility). This allowed us to discard a dozen discharge events. 
{Note that the RPWS measurements later in the mission were no more affected by the MIMI/INCA discharges issue because these were significantly diminished after the negative voltage mode was turned off on day 13 of 2001.}

\subsection{Interpretation of the voltage spectra}
Let us now check whether the dust signatures  identified in Figure \ref{RPWS_Spectro} are consistent with impacts of the interplanetary nanograins picked-up by the solar wind, identified by  \cite{MeyerVernet2009b}. 

The  dust signatures are observed only in monopole mode and are similar on the three monopole antennas. This indicates that the dust is detected via charge  recollection by the spacecraft (since the spacecraft is positively charged in the solar wind, these charges are electrons.) Such a detection mechanism is similar to that by which nanodust was detected near Jupiter by Cassini/RPWS \citep{MeyerVernet2009a}.

The maximum voltage $\delta V$ produced by a dust grain impact on the spacecraft and measured by each monopole antenna is:
 \begin{linenomath*}
\begin{equation}
\delta V \simeq \Gamma Q/C
\label{deltaV}
\end{equation}
 \end{linenomath*}
where $Q$ is the charge released, $C$ is the spacecraft capacitance (200 pF) and $\Gamma$ is the antenna gain $\simeq 0.4$ \citep{Gurnett2004}.
The impact generated charge Q strongly depends on the grain speed and can be approximated by  \citep{McBride1999}: 
 \begin{linenomath*}
\begin{equation} 
Q=0.7 m v^{3.5} 
\end{equation}
 \end{linenomath*}
with $Q$ (Cb), the grain mass $m$ (kg) and {$v$ (km/s)}. 
Nano-sized grains picked-up by the solar wind can be accelerated up to a speed $\simeq$ 300 km/s near 1 AU provided that their mass is smaller than about $5 \times 10^{-20}$ kg  \citep{Mann2010}. 

We determine the flux  of nano grains by equating the {theoretical Equation} (\ref{Eqn_Dust2}) and measured power spectra $V_f^2$  at $f=10$ kHz  since at and above this frequency, the spectrum varies approximately as $f^{-4}$. 
The theoretical power spectral density $V_f^2$ induced by a distribution of grains with flux $F(m)$ is:
 \begin{linenomath*}
\begin{equation}
V_f^2\simeq S \int_{m_{min}}^{m_{max}} dm \frac{dF(m)}{dm} V_{fi}^2(m) 
 \label{Eqn_Vf_tot}
 \end{equation}
  \end{linenomath*}
where $S$ is the effective spacecraft surface area ($\simeq$ 15 m$^2$), and $m_{min}$ and  $m_{max}$ are respectively the grain minimum and maximum detected mass. We assume  that: 
\newline - the cumulative flux is of the form 
 \begin{linenomath*}
\begin{equation}
F(m) = F_0m^{-5/6}
\label{F_PowerLaw}
\end{equation}
 \end{linenomath*}
according to the interplanetary dust distribution model  by \cite{Grun1985} {for masses below 10$^{-18}$ kg}.
\newline - the spectral density induced by one grain $V_{fi}^2$ is defined by equation (\ref{Eqn_Dust2}) where $\delta V$ is given by  equation (\ref{deltaV}) so that 
{ \begin{linenomath*}
\begin{equation}
V_f^2\propto F_0 m_{max}^{7/6}/\omega^4
\label{Vf2Mmax}
\end{equation}
 \end{linenomath*}
since m$_{min}$ is much lower than m$_{max}$.}

{The maximum mass $m_{max}$ is the smallest between i) the mass $m_0$ ($\simeq$5$\times$10$^{-20}$kg) above which the grains are no longer significantly accelerated by the solar wind and ii) the greatest mass that can impact the spacecraft during the integration time $\Delta t \simeq 0.15$ s, i.e. such that $F(m_{max})S \Delta t \simeq 1/e$. Hence 
 \begin{linenomath*}
\begin{equation}
m_{max}= \rm{min}[\it{m_0,(eS\Delta t F_0)^{6/5}}]
\label{Mmax}
\end{equation}
 \end{linenomath*}
 }
Figure \ref{F0_vs_time} displays the electric power spectral density measured by RPWS/HFR between 3 and 200 kHz (Top panel) and the normalized flux $F_0$ (Bottom panel) deduced  for the time period from August 20 (day 232) to  September 15 (day 258) of 1999.

In order to eliminate the contributions of discharges, radio and plasma emissions, and plasma quasi-thermal noise, we have carefully selected the times  when:
\newline 1) the receiver does not display signatures due to MIMI/INCA discharges, according to section 3.1 (red vertical lines on Figure \ref{F0_vs_time}b indicate the times when discharges were identified),
\newline  2) the average index of the power law adjusted to the measured spectrum lies between $-$2.2 and $-$4, 
\newline 3) the measurements of the three monopoles (u, v, w) are similar {(within 50$\%$)} when the instrument is in direction finding mode  \citep{MeyerVernet2009a}, and
\newline  (4) the ratio between the measurement of monopole w and the dipole is greater  than a factor of 3. As discussed in Section 2, conditions 3) and 4) enable us to eliminate contributions from plasma wave electric fields.

Figure \ref{F0_vs_time}b shows that most of the low-frequency electric power density enhancements are identified as dust signatures according to our criteria. The normalized flux values $F_0$ we derived oscillate from 10$^{-19}$ and 10$^{-17}$ m$^{-2}$s$^{-1}$kg$^{5/6}$ during the whole time period. We note that the strong criteria may underestimate the number of dust detections.  {Relaxing this selection criteria would increase by 35\% the number of detections while including outliers (i.e. $F_0$ bigger than 10$^{-16}m^{-2}s^{-1}kg^{5/6}$). We note then that our data selection does not change the order of magnitude of the results.}
      
Figure \ref{Cumulative_Flux} displays the cumulative flux (\ref{F_PowerLaw}) (in yellow) calculated from the RPWS/HFR  during the time interval displayed in Figure \ref{RPWS_Spectro}a. The minimum, maximum and average flux are displayed in red.  Cumulative flux derived from the low frequency receiver LFR \citep{MeyerVernet2009b} and the waveform TDS \citep{Zaslavsky2012} of STEREO/WAVES instrument are superimposed with black crosses as well as the interplanetary dust distribution model \citep{Grun1985} (green curve). As noted above, since the values displayed correspond to selected data, the minimum flux indicated is not the actual minimum flux during the period considered, but the minimum flux when nanodust signatures are observed without any ambiguity, which represents a small fraction of this period. 
With this proviso, the Cassini results are consistent with the previous interplanetary nanodust detection.
{The flux is also consistent with the most likely value of the flux of particles of mass greater than 10$^{-20}$ kg derived from the statistical survey from \cite{LeChat2013} based on five years (from 2007 to 2011) of STEREO/WAVES/LFR data (Figure 5 in that paper).}

\section{Discussion} 

{We analyzed the high frequency radio data measured onboard Cassini spacecraft during its cruise phase in the solar wind near 1 AU and identified a bursty low-frequency noise which is consistent with the impacts of nano-sized grains on the spacecraft surface. The present analysis, based on Cassini/RPWS data, supports and confirms the presence of the nanograins in the solar wind near 1 AU as previously identified by \cite{MeyerVernet2009b} and \cite{Zaslavsky2012} from STEREO/WAVES data analysis and their bursty nature. 

{The observed variability with burst-like features is not surprising and was also observed on STEREO \citep{MeyerVernet2009b,Zaslavsky2012,LeChat2013}. First of all, because of {Equations (\ref{Vf2Mmax}) and (\ref{Mmax})}, the power density varies very fast with $F_0$. Indeed, $V_f^2 \propto F_0^{12/5}$ when $m_{max}$ is determined by the right-hand side term in the bracket of equation (\ref{Mmax}), which occurs most of the time. Hence, a small variation in flux produces a large variation in power spectral density. Because of the short transit time from the inner solar system to 1 AU, the intrinsic flux variability may be produced by variations in nanodust trapping near the Sun  due to magnetic field changes in the inner solar system \citep{Czechowski2012}, as well as to changes in dust collisional rates \citep{Mann2012}. Furthermore,  the Lorentz force driving nanodust trajectories varies in permanence at virtually all scales due to variations in both solar wind speed and magnetic field, producing variability in nanodust flux and arrival direction \citep{Mann2012, Juhasz2013}.}

Although not designed for dust detection, the high frequency receiver of Cassini/RPWS has demonstrated its capability to sound nanodust in the interplanetary medium.}
{Similar signatures in a region close to the asteroid belt at 2.7 AU are now under study.}

%
%

\begin{acknowledgments}
The data are from the RPWS/HFR receiver of Cassini and are hosted in LESIA, Observatory of Paris-Meudon. The research at LESIA (Observatory of Paris) is supported by the CNES (Centre National d'Etudes Spatiales) agency. The research at the University of Iowa was supported by NASA through contract 1415150 with the Jet Propulsion Laboratory.
 
 
\end{acknowledgments}

\end{article}

 \setcounter{figure}{0}
 
\begin{figure}[htbp]
\includegraphics[width=42pc]{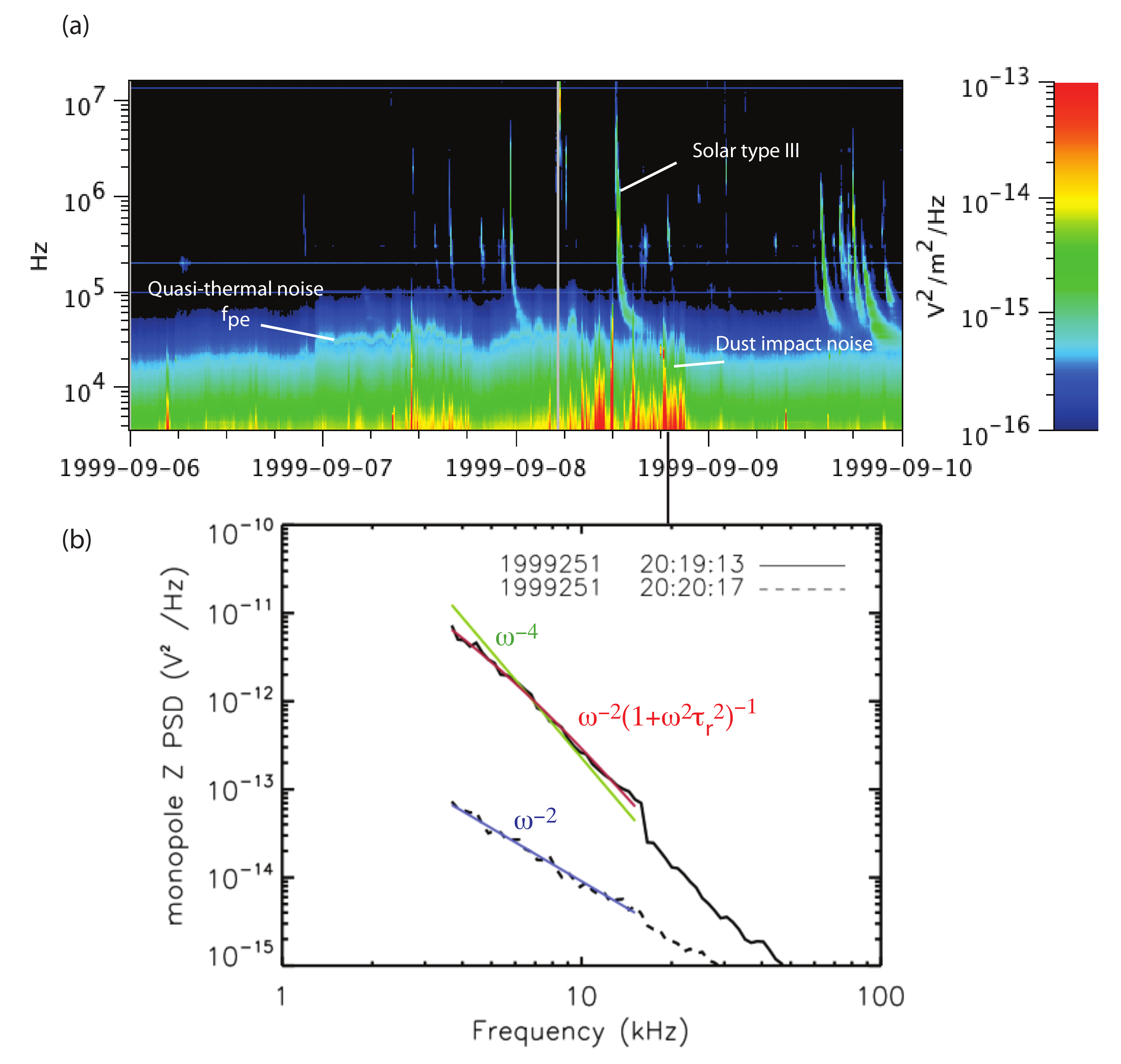}
\caption{Panel a: RPWS/HFR dynamic spectrum in electric power spectral density ($V^2/m^2/Hz$) between day 249 (September 6) of 1999 and day 253 (September 10) of 1999 with the antenna z in monopole mode. Panel b: Voltage power spectral density spectra at 20:19UT and 20:20UT on day 251 (September 8) of 1999, in solid and dashed lines respectively. Three power spectral density models are superimposed: 1) V$_f^2$ $\propto\omega^{-4}$ (in green), 2) V$_f^2$ $\propto\omega^{-2}$ (in blue) and 3), V$_f^2$ $\propto\omega^{-2}(1+\omega^2\tau_r^2)^{-1}$ (in red).}
\label{RPWS_Spectro}
\end{figure}

\begin{figure}[htbp]
\includegraphics[width=42pc]{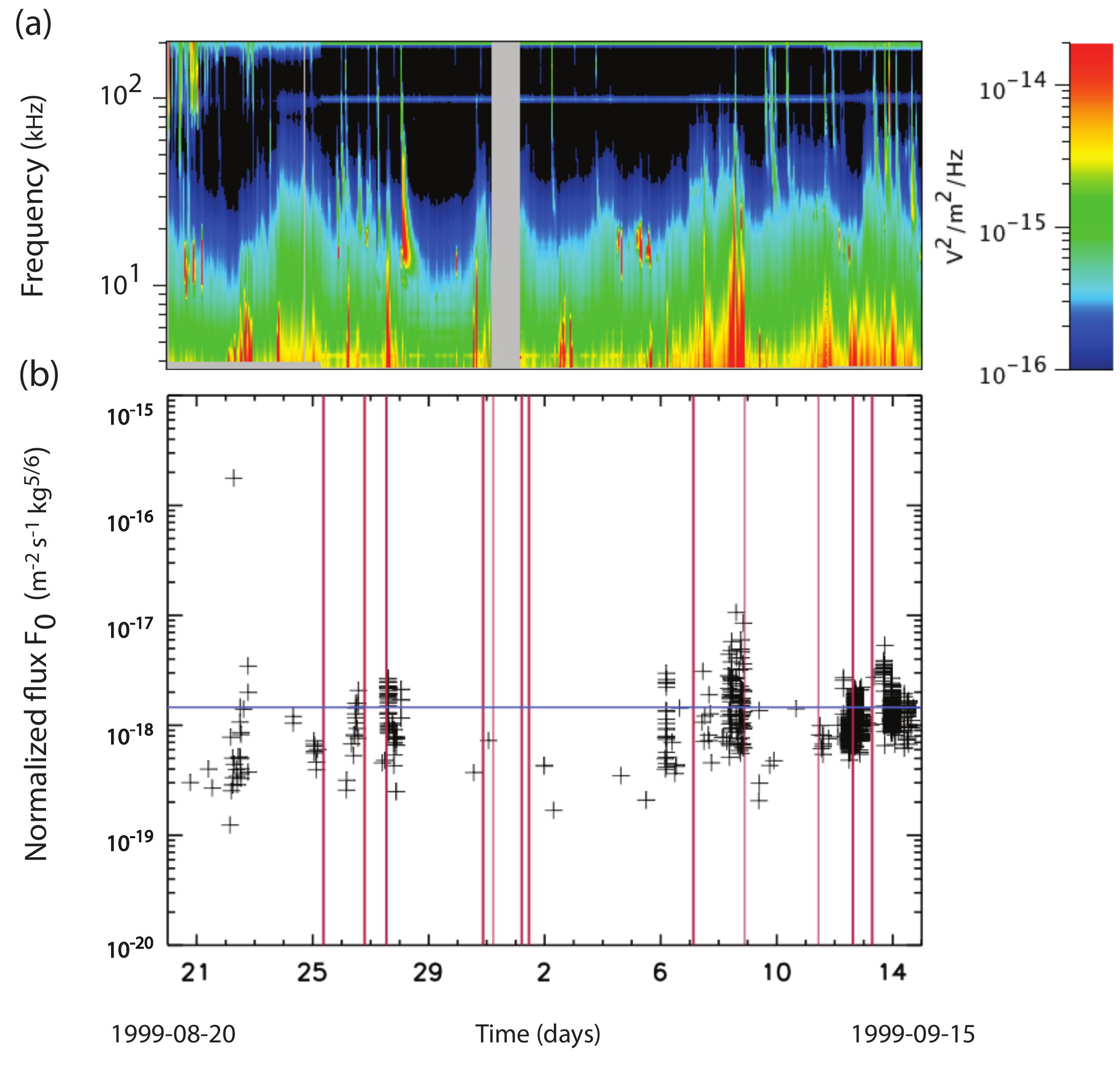}
\caption{Panel a: RPWS/HFR dynamic spectrum in electric power spectral density ($V^2/m^2/Hz$) below 200 kHz, between day 232 (August 20) of 1999 and day 258 (September 9) of 1999. Panel b: Coefficient $F_0$ determined from our analysis in function of time. The horizontal blue line represents the average value of $F_0$ calculated over the whole time interval. The red vertical lines indicate the times when MIMI/INCA discharges were identified.}
\label{F0_vs_time}
\end{figure}

\begin{figure}[htbp]
\includegraphics[width=42pc]{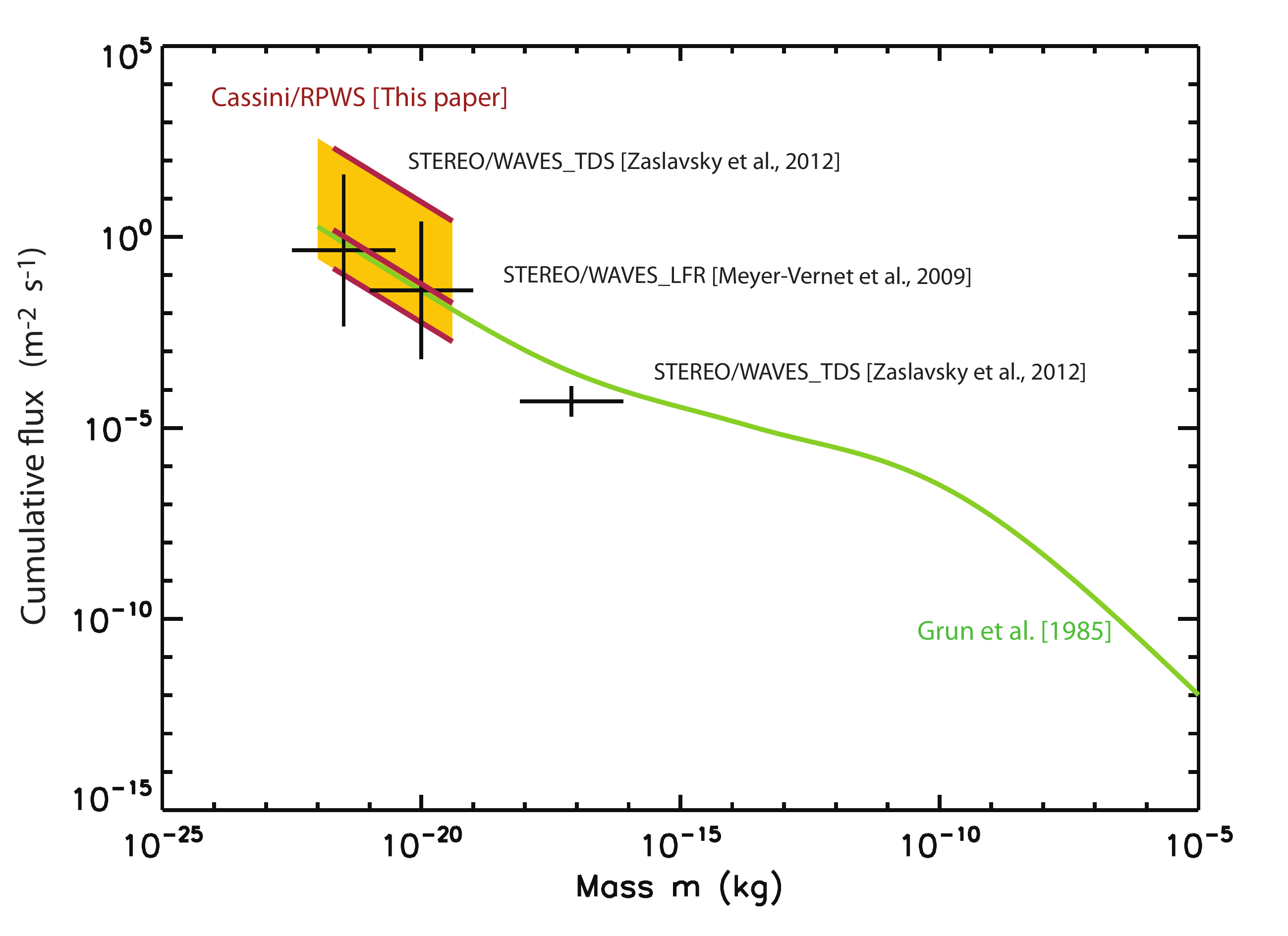}
\caption{Cumulative flux (m$^{-2}$s$^{-1}$) in function of the mass (kg). Our results using the form $F=F_0m^{-5/6}$ is displayed with a thick red line. 
STEREO results from \cite{MeyerVernet2009b,Zaslavsky2012} are displayed with black crosses. \cite{Grun1985} model is displayed with a green heavy line.}
\label{Cumulative_Flux}
\end{figure}

\end{document}